%% file: aimc2025.tex
\documentclass{article}

\usepackage{aimc2025}

\usepackage[utf8]{inputenc} 
\usepackage[T1]{fontenc}    
\usepackage{hyperref}       
\hypersetup{
  colorlinks   = true, 
  urlcolor     = blue, 
  linkcolor    = black, 
  citecolor   = black 
}
\usepackage{url}            
\usepackage{booktabs}       
\usepackage{amsfonts}       
\usepackage{nicefrac}       
\usepackage{microtype}      
\usepackage{graphicx}       
\usepackage{algorithm}
\usepackage{algpseudocode}
\usepackage{amsmath}

\title{Incorporating Structure and Chord Constraints in Symbolic Transformer-based Melodic Harmonization}

\author{%
  Maximos Kaliakatsos-Papakostas, Konstantinos Soiledis, \\ \textbf{Konstantinos-Theodoros Tsamis,
  Dimos Makris}\\
  Department of Music Technology and Acoustics\\
  Hellenic Mediterranean University, Greece, and \\
  Archimedes, Athena RC \\
  \texttt{\{maximoskp, ksoiledis, ktsamis, dimakr\}@hmu.gr} \\
  \And
  Vassilis Katsouros \\
  Institute of Language and Speech Processing\\
  and Archimedes, Athena RC, Greece\\
  \texttt{vsk@athenarc.gr} \\
  \And
  Emilios Cambouropoulos \\
  School of Music Studies\\
  Aristotle University of Thessaloniki, \\
  Greece\\
  \texttt{emilios@mus.auth.gr} \\
}

\begin{document}

\maketitle

\begin{abstract}
  Transformer architectures offer significant advantages regarding the generation of symbolic music; their capabilities for incorporating user preferences toward what they generate is being studied under many aspects. This paper studies the inclusion of predefined chord constraints in melodic harmonization, i.e., where a desired chord at a specific location is provided along with the melody as inputs and the autoregressive transformer model needs to incorporate the chord in the harmonization that it generates. The peculiarities of involving such constraints is discussed and an algorithm is proposed for tackling this task. This algorithm is called B* and it combines aspects of beam search and A* along with backtracking to force pretrained transformers to satisfy the chord constraints, at the correct onset position within the correct bar. The algorithm is brute-force and has exponential complexity in the worst case; however, this paper is a first attempt to highlight the difficulties of the problem and proposes an algorithm that offers many possibilities for improvements since it accommodates the involvement of heuristics.
\end{abstract}

\input{0_intro}
\input{1_method}
\input{2_results}
\input{3_conclusions}

\begin{ack}
This work has been partially supported by project MIS 5154714 of the National Recovery and Resilience
Plan Greece 2.0 funded by the European Union under the NextGenerationEU Program.
\end{ack}

\section*{Ethics statement}

The presented work involves methods for supporting human control on AI systems and moves toward the direction of better understanding and manipulating AI system decisions. No subjective experiments were conducted. We acknowledge that the proposed algorithm involves a significant number of model calls and this could have a significant energy impact, however, in the context of the presented study the models were small in size the impact was minimal.



\bibliographystyle{plainnat}
\bibliography{references}  

\end{document}

%% file: 0_intro.tex
\section{Introduction}\label{sec:intro}

Melody harmonization is the task of assigning chords at the proper positions in a given melody. This task is very valuable for examining properties of symbolic sequence generation systems, since it requires understanding global context (e.g., mode and tonality), non-local relations (e.g., melody-harmony compatibility and structural repetitions) and local context (e.g., how successive chords interact). Notable research has focused on the employment of autoregressive generation methods for this task, including transformers~\citep{NIPS2017_3f5ee243}. Such models take the melody as input and autoregressively generate the harmony by completing the harmonic sequence one token at a time. Applying chord constraints, i.e., ensuring that the generated harmony will include a specific chord in a specific location of a specific bar, is not trivial in such autoregressive models, since they are not capable of considering future conditions while they generate: their input incorporates only the melody and the harmony tokens that have been generated so far. Additionally, including chord constraints is a way to incorporate human input when it comes to melodic harmonization.

Regarding the current state of the literature in chord-to-chord generation~\citep{agosti2020transformer} and melodic harmonization, the most common representation of harmony is through chord symbols, i.e., where each chord is represented by its own token. Deep learning methods that are based on BiLSTM architectures and a small number of symbols, e.g., 24~\citep{costa2023neural} or 48~\citep{yeh2021automatic, chen2021surprisenet} have been studied, as well as models with larger vocabularies of chord symbols~\citep{sun2021melody, zeng2021automatic}, and up to 1462 chord symbols~\citep{wu2024generating}. Such tokenization approaches have been popular with transformer architectures as well. Examples of various transformer architectures for harmony generation were presented in~\citep{rhyu2022translating, huang2024emotion, wu2024melodyt5} and for four-part harmonization generation in~\citep{zhouchoir}.

Another approach to tokenizing harmony is by breaking down chords into their constituent pitch classes or even octave-informed pitches~\citep{cholakov2018ai}. Such chord ``spelling'' approaches have been employed in LSTM / GRU-based approaches~\citep{lim2017chord, wang2020learning, ji2023rl, ji2023emotion} and Transformer-based systems~\citep{huang2018musictransformer}. The aim of this paper is to study chord constraint mechanisms across both tokenization paradigms, i.e., where the constraint chord is expressed either as a unique token of the chord symbol, or a succession of tokens that correspond to the tokens of the pitch classes that constitute the chord. In both approaches, the token that indicates the position of the chord directly precedes the chord token or the succession of chord tokens, while the position-and-chord subsequence of tokens needs to be properly placed between
the correct bar tokens. There are other possible ``intermediate'' tokenization approaches, e.g., where chords are tokenized in root-type pairs~\citep{dalmazzo2023chordinator}; in the current work, however, we are focusing on the two ``extremes'': chords as single tokens and chords as collections of pitch class tokens.

Although constraint-based generation is well-studied in natural language processing~\citep{liang2024controllable}, incorporating hard constraints in melody-driven harmony generation presents unique challenges. The NLP literature on hard constraints generally identifies three classes of methods. The first class involve generation using encoder-only models~\citep{zhang2020pointer, he2021parallel, hsieh2021enconter, jonason2024steer} through successive insertion of tokens starting from a constraint sequence; a similar approach was followed by Coconet using convolutional neural networks~\citep{huang2019counterpoint}. The second class involves autoregressive decoder architectures and examines manipulation strategies on the continuous distributions of logits before they materialize in discrete tokens, based
on energy-related optimization~\citep{qin2022cold, kumar2022gradient}. These methods, however, are not applicable to the problem at hand because: (a) they consider that the length of the sequence is determined beforehand and remains fixed throughout generation\footnote{Changing a chord in a given harmonization to a desired constraint chord is very likely to lead to the addition or removal or other chords, rather than simply changing other chords, a fact that changes the overall length of the sequence.} and (b) it is not straightforward to describe hard constraints in a differentiable way that is appropriate for the formulation of the problem at hand.

The class of methods that are more relevant to the paper at hand, assume pretrained autoregressive generation transformers (GPT-style)~\citep{hokamp2017lexically, post2018fast, pascual2020directed, lu-etal-2022-neurologic}. These methods work under the assumption that the necessary word constraints will be naturally selectable at some point during the generation process. This assumption may be applicable to text-related tasks. For example, in language translation the expected meaning of a word to be translated can be expressed through multiple words. Constraining generation to a specific word among those will not lead to unresolvable conflicts between contradicting semantics, since the semantic space has been prescribed by the sentence to be translated. Additionally, the order in which words appear is usually not detrimental to the syntax and the meaning of the sentence. For example, the phrases ``I was happy with the result'' and ``the result made me happy'' have almost the same meaning, but the word ``happy'' appears in different positions within the sentence.

In harmony, however, applying a chord constraint requires that there are no conflicting chords (semantic integrity) and that the chord is placed at a very specific location within the tokens, i.e., placed properly between the tokens that indicate bar position and after a specific chord position token. Therefore, incorporating chord constraints at specific locations in autoregressive harmony generation systems can \textit{not} be performed by algorithms that work under these two assumptions: (a) the constraint chord will ``naturally'' occur at some point (the desired chord constraint might be outside the expected context described by the melody, e.g., a secondary dominant) and (b) that the algorithm can try to insert the constraint at any point during generation (since the position it needs to be placed is very specific and the exact place of the token within the generated sequence cannot be known beforehand). To further highlight how the constraints cannot simply occur in a ``natural'' way, a special case of the infilling algorithm proposed in~\citep{donahue2020enabling} is examined, that shows the inability of pretrained melody harmonization transformers to comply with constraints when they are provided as parts of the input.

Therefore, a system that performs autoregressive, only forward-looking harmony generation will inevitably need to try out multiple harmony paths, canceling paths that ultimately failed to reach the constraint, until it reaches one path that more ``naturally'' places the constraint chord at the proper position. The proposed algorithm does exactly that: it combines ideas from beam search and the A* algorithm to try out the optimal or more promising paths first, before it starts backtracking and trying out optimal and promising alternatives. The backtracking stage aims to find solutions that are optimal in terms of overall structure and, therefore, manipulates parents of nodes that fail to produce children that satisfy the constraints. Because of the recursive nature of the algorithm and due to the fact that it combines components from beam search and A*, we name this algorithm \textit{B*}. In the worst case, the complexity of the algorithm is exponential. Even though harmony generation can be limited to a few tokens (e.g., generating 8 or 4 bars at a time), this complexity makes this algorithm unsuitable for real-world applications. This paper, however, offers the following contributions:
\begin{enumerate}
    \item We introduce the B* algorithm to enforce hard chord constraints during generation. This
    design was motivated by the observation that standard autoregressive decoding frequently
    fails to include specified chords in the correct position, especially when the requested
    harmony contradicts the model’s learned statistical priors. For example, a left-to-right
    Transformer might ignore a user-specified secondary dominant if it doesn’t fit the model’s
    internal expectations. B* overcomes this limitation by exploring multiple hypotheses and
    performing backtracking as needed to guarantee that the constraint is satisfied, even under
    unlikely or out-of-distribution conditions.
    \item We provide an empirical analysis of B*’s complexity in practice to gauge its feasibility. This analysis shows how often the worst-case (exponential) behavior occurs and identifies typical runtimes, validating the practicality of our approach within certain limits.
    \item To contrast with B*, we evaluate a soft-constraint formulation, where suggested chords are simply appended to the input as special tokens. This experiment demonstrates why a naive approach fails: the model often ignores user-provided chord suggestions when they conflict with learned melodic–harmonic priors. This analysis emphasizes the necessity of algorithms like B*, as pretrained Transformers lack explicit mechanisms for enforcing hard musical conditions.
\end{enumerate}
The code of the paper can be found online\footnote{\url{https://github.com/NeuraLLMuse/BStarConstraintHarmonization}}.

%% file: 1_method.tex
\section{Autoregressive generation with constraints}\label{sec:method}

This section describes all the methodological components that were examined for melodic harmonization. These include melody and harmony tokenization, the architectures of the examined model, the data and the B* algorithm description.

\subsection{Tokenization}\label{subsec:tokenization}

Two chord tokenization methods are employed that express lead sheet music information, i.e., chords and melody notes. These representations include some well-established principles for general purpose tokenization of symbolic music (e.g., tokens for bars and for position of notes and chords). The difference between these methods is that one views chord symbols as independent tokens while the other breaks down chord information to its constituent pitch classes.

\subsubsection{Melody tokenization}\label{subsubsec:melody}

Since the task is melodic harmonization, only basic information of the melody is tokenized. Both examined tokenization methods include tokens that indicate sequence start, end, mask, unknown and padding; they also include a common melody tokenization method with special tokens indicating the occurrence of a bar (\texttt{<bar>}), of a rest (\texttt{<rest>}), the onset time (\texttt{position\_BxSD}) and the MIDI pitch (\texttt{P:X}) of notes. Regarding the onset position, it is provided in the form \texttt{BxSD}, where \texttt{B} shows the beat within the bar and \texttt{SD} the beat subdivision quantized in eight parts\footnote{The following beat subdivisions are considered: $\{0, 1/6, 1/4, 1/4, 1/2, 2/3, 3/4, 5/6\}$, corresponding to
resolution of up to 16th note triplets.}. Additionally, melody tokenization includes the time signature of the entire piece; a single time signature is considered at this point, since the dataset comes separated in parts with a single time signature. The time signature token is in the form \texttt{ts\_NxD}, where \texttt{N} denotes the numerator and \texttt{D} the denominator of the time signature. For the purposes of the employed dataset, numerator values from 1 to 10 and denominator values 4 or 8 were employed.

\subsubsection{Harmony tokenization}\label{subsubsec:harmony}

We employ two harmony encoding strategies – one treating each chord symbol as a single token, and another breaking chords into multiple pitch-class tokens – to study the trade-off between a large discrete chord vocabulary and a fine-grained representation. The intuition is that a chord-symbol tokenizer (CS) provides a holistic chord representation that might make enforcing specific chords easier, whereas a pitch-class tokenizer (PC) offers more flexibility and detail (spelling out chords note-by-note) at the cost of longer sequences. By comparing them, we can assess how token granularity affects constraint incorporation.

Both examined harmony tokenization methods follow a unified baseline tokenization scheme that borrows tokens from melody tokenization for unknown, padding and mask and position tokens. All harmony tokenizers also include a special token that signifies the start of harmony in the sequence (\texttt{<h>}). The two harmony tokenizers are different in how they represent chords:
\begin{enumerate}
    \item \texttt{ChordSymbolTokenizer}: Each token describes the symbol of the chord as it appears on the lead sheet, e.g., \texttt{C:maj7}.
    \item \texttt{PitchClassTokenizer}: No chord symbols are used, instead, the pitch classes of the chord are spelled out, e.g., the tokens that describe a Cmaj7 chord are \texttt{chord\_pc\_0}, \texttt{chord\_pc\_4}, \texttt{chord\_pc\_7} and \texttt{chord\_pc\_11}.
\end{enumerate}
Table~\ref{tab:examples} shows an example of two bars being tokenized with the examined tokenization methods.

\begin{table}
    \centering
    \begin{tabular}{p{0.25\linewidth}p{0.7\linewidth}}
        \textbf{Tokenizer} & \textbf{Example} \\ \hline \hline
        \texttt{ChordSymbolTokenizer} & \texttt{<h> <bar> position\_0x00 C:maj <bar> position\_0x00 E:min position\_2x00 G:maj} \\ \hline
        \texttt{PitchClassTokenizer} & \texttt{<h> <bar> position\_0x00 chord\_pc\_0 chord\_pc\_4 chord\_pc\_7 <bar> position\_0x00 chord\_pc\_4 chord\_pc\_7 chord\_pc\_11 position\_2x00 chord\_pc\_7 chord\_pc\_11 chord\_pc\_2} \\ \hline
    \end{tabular}
    \caption{Tokenized examples of the first two bars in a harmonic sequence using the two examined tokenizers.}
    \label{tab:examples}
\end{table}

\subsubsection{Structure and chord constraint along with the melody}\label{subsubsec:structure}

To incorporate a user-specified chord as a ``soft'' constraint, we append in the model input a sequence of tokens after the melody that encode the song’s bar structure and the desired chord. For example, say the user wants a C major chord on the first beat of bar 3. We mark the end of the melody with a special token \texttt{</m>}, then list the bar tokens and filler tokens ~\citep{donahue2020enabling} for each bar until the constraint: \texttt{<bar>} \texttt{<fill>} \texttt{<bar>} \texttt{<fill>} ... up to bar 3, where we place \texttt{position\_1x00} \texttt{C:maj} to specify a C major at beat 1 of that bar. After that, we continue with \texttt{<fill>} tokens for any remaining structure, and finally the \texttt{<h>} token ends the input to the model, indicating the start of harmony (output) generation. During training, the model learns to associate this structured prompt with a harmonic sequence that indeed contains the specified chord in bar 3. Essentially, \texttt{<fill>} tokens tell the model ``please generate a chord here'', while an actual chord token in the sequence tells the model ``use this specific chord at this position''.

\begin{verbatim}
    [MELODY TOKENS] </m> <bar> <fill> <bar> <fill> <bar> <fill> [...]
    <bar> <fill> position_1x00 C:maj <fill> <bar> <fill> [...] <h>
\end{verbatim}
During training, the harmonic token sequences that are learned correspond exactly to the structure indicated by the respective input sequences between the \texttt{</m>} and \texttt{<h>} tokens, including the desired chord at the specific position. The aim of this protocol is to examine whether the inclusion of this extension in the input tokens improves the overall output of the models without it and to what extent it leads to the incorporation of the chord suggestions.

\subsubsection{Transformer architectures and data}\label{subsubsec:transformers}

We evaluate both an encoder–decoder architecture (BART) and a decoder-only architecture (GPT-2) for harmonization to investigate how each handles chord constraints. BART encodes the full melody context, potentially allowing a richer understanding of future chords, whereas GPT-2 generates harmony as a continuation, mimicking how a musician might continue a sequence. Comparing these lets us see which paradigm better integrates user constraints.

Two basic transformer-based architectures are examined for generating melodic harmonization:
\begin{description}
    \item[Encoder-decoder] melody harmonization performed by BART~\citep{lewis2019bart}, where the melody is used as input to the encoder and the harmony is generated autoregressively in the decoder until a stopping criterion (end-of-sentence token is generated or maximum number of tokens is achieved).
    \item[Decoder-only] melody harmonization performed by GPT2~\citep{radford2019gpt2}, which autoregressively generates the harmony tokens given an initial prompt that comprises melody tokens and the structure and constraint tokens when they are used.
\end{description}

For both approaches, there is a maximum limit of 512 tokens in the melody and another 512 in the harmony part, but both melody and harmony need to include the same number of bars. To this end, when tokenizing a lead sheet if either the melody or the harmony component surpasses the maximum tokens limit, the tokens corresponding to the final bar of both the melody and the harmony are removed; this process is repeated until the number of tokens in both melody and harmony parts is below the 512-token limit. This process ensures that the melody and harmony parts include the same number of bars during training.

The dataset comprises the lead sheet version of 17,476 files from the Hook Theory dataset, which was initially compiled in~\citep{yeh2021automatic} and primarily contains popular Western music, especially pop, rock, and related mainstream genres. These files were converted into a proper lead sheet format and stored as musicXML files. The collected pieces are composed in different keys, with some keys being more prevalent than others. To remove key-related biases in the dataset, all pieces were transposed to a key with no sharps or flats, i.e., major-like mode pieces were transposed to C major and minor-like mode pieces to the key of A minor respectively. The tonality of each piece was computed using the Krumhansl key-finding algorithm, which determines the key of a musical piece by correlating its pitch-class distribution with empirically derived key profiles~\citep{krumhansl2001cognitive}. Other works transpose music pieces to C major and C minor keys~\citep{rhyu2022translating} or use functional harmony~\citep{huang2024emotion}, which is equivalent to the C major - C minor transposition. All pieces were transposed to a key with no sharps or flats (C major for pieces originally in major mode, A minor for minor-mode pieces) to normalize the data and eliminate key-specific biases. This relative major/minor transposition~\citep{hahn2024senthymnent} leverages the shared pitch collection of C major and A minor. It creates a unified context where chord–melody relationships are comparable across pieces, since C major and A minor share the same pitches. Transposing to C major and A minor, as in~\citep{hahn2024senthymnent}, however, leverages the relative relation between these two tonalities, creating a shared context where the relations between all chords and the respective melody notes are the same. A random training - testing/validation set split of 90\%-10\% was used leading to a total of 15956 pieces for training and 1520 for testing and validation during training.

\subsection{The \textit{B*} algorithm}\label{subsec:bstar}

As shown later in the Results section, simply incorporating the chord constraint in the input sequence does not guarantee the constraint will be satisfied. To this end, the B* algorithm is proposed that incorporates components of the beam search algorithm~\citep{lowerre1976harpy} and the A* algorithm~\citep{hart1968formal}, which ensures the inclusion of the chord constraint using a partially greedy search strategy to go through the most probable paths first and fall back to less probable if necessary. This combination of the two algorithms is using a restricted beam-style best-first approach to try to reach a possible solution that satisfies the constraints quickly, but it does keep all possible solutions in an open set, which is then used, if no solution was found quickly in the beam set of sequences, for backtracking to move through unvisited token sequences. A scoring function selects possible continuation of the sequences that are more ``mature'', in a sense that they are optimal in terms of model log likelihood and closer to the length that the constraint is expected, thus enhancing the possibility to produce a good result more quickly. Additional functions act as heuristics to check for the consistency of the generated solutions at each step. Algorithm~\ref{alg:astar} is an abstract overview of the examined approach.

\begin{algorithm}
\caption{Decoding with Chord Constraint}\label{alg:astar}
\begin{algorithmic}[1]
\State \textbf{Input:} Model $M$ and tokenizer $T$, melody tokens $m$, constraint specification $C$, a beam width value $b$, en expansion value $k$
\State \textbf{Output:} Harmony sequence $h$ satisfying chord constraint $C$

\State Initialize an empty beam queue $B$ and a priority queue $Q$; $Q$ initially contains $x$

\While{queue $Q$ is not empty}
    \For {i in min\{$b$, |Q|\}}
        \State Pop the sequence / node $s$ with highest score from $Q$
        \If{$s$ ends with EOS and satisfies $C$}
            \State \label{step:found} Set $s$ as the desired sequence and \textbf{break}
        \ElsIf{sequence length exceeds model maximum}
            \State \textbf{continue}
        \EndIf
        
        \State Generate top-$k$ token predictions from model $M$ at current node
        \For{each top-$k$ token $t$}
            \State Extend current sequence with $t$
            \If{resulting sequence is well-formed and satisfies constraint}
                \State Compute score combining log-probability
                \State Add resulting node to queue $B$
            \EndIf
        \EndFor
    
        \If{no valid children found}
            \State Backtrack to parent node and retry with unvisited options
        \EndIf
    \EndFor
    \If{desired sequence found}
        \State \textbf{break}
    \EndIf
    \State Pop all nodes of $B$ and push them into $Q$
\EndWhile

\State \Return Desired sequence (see step~\ref{step:found})

\end{algorithmic}
\end{algorithm}

This algorithm receives a pretrained model and its tokenizer, the melody tokens, the constraints, a beam width value $b$ and a look ahead value $k$; in the current implementation the constraints are provided as tokens in the form shown earlier in Section~\ref{subsubsec:structure} using the tokens of the constraint chord and chord position tokens placed properly among the \texttt{<bar>} and \texttt{<fill>} tokens. The algorithm produces harmony tokens that satisfy the chord constraints.

Regarding the involvement of the A* algorithm, the typical version traverses nodes on a weighted graph and tries to find the optimal path from one to another selected nodes. In the proposed algorithm, the nodes are sequences of tokens (not single tokens); transitioning from one node to the next means appending a token at the sequence that represents the former node. Each possible new token that can be appended to an existing node, leads to a possible new node. The initial node (token sequence) is an empty harmonic sequence to be filled by the input model. In the case of GPT2 model the initial $x$ is the melody tokens with the ``start harmony'' token (\texttt{<h>}) appended at the end. In the BART model case, the initial $x$ is simply \texttt{<h>}.

At each step, the top-scored (scoring is discussed later) $b$ ``parent'' sequences $s$ in $Q$ are expanded by running the model for one prediction, generating ``children'' nodes that are one token larger; the top-$k$ predicted tokens, which are used to construct the next $k$ possible nodes (sequence expansions / continuations). These top-$k$ possible continuations / nodes generated by expanding for one token each of the $b$ initial sequences are checked in terms of token consistency (using some basic heuristics of well-formedness, different for each tokenizer) and constraint satisfaction. Constraint satisfaction is only examined for the sequences that have reached the bar of the constraint and have reached or exceeded the position of the constraint chord within this bar. The sequence continuations that are consistent and satisfy the chord constraint are put in a list $B$ that holds the sequences / nodes that are currently examined (running ``beam''). After the first $b$ sequences in $Q$ have been expanded and scored, they are placed in $B$. Up to this point, the algorithm acts as the typical beam search algorithm.

The A* algorithm components come input play at this point, since the components in $B$ are popped out ($B$ gets emptied) and inserted into $Q$, which acts as the open set. The process starts over, but by selecting the best rated $b$ values from $Q$, which holds the entire open set of nodes that are consistent and satisfy the constraint so far. Therefore, up to this point, a beam-search pace is maintained (since only $b$ sequences are considered at each time), but all other sequences remain ``dormant'' until they are needed.

If all children nodes of a currently examined beam are blocked, i.e., fail to satisfy the constraint, then a backtracking process begins that expands the parent of the currently selected node; each node ``remembers'' its children nodes (specifically, next token of each of their children) and every parent expansion involves only the tok-$k$ unvisited token continuations. These new children are then inserted in to the beam set $B$ and subsequently into the open set $Q$. Whenever a child node is a complete sequence (ending with the end-of-sequence token) that satisfies the constraint, execution ends successfully and this child is the desired result. 

Calculating the score of each node / sequence is a crucial component in the implementation as it impacts not only the efficiency of the algorithm, but also the quality of the results. A straightforward approach would be to score each sequence by summing all the log probability contributions of each token within the sequence and this makes sense, since longer sequences are expected to be closer to completion and, therefore, closer to a successful termination of the algorithm. However, in cases of chord constraints that are considered as low probability for the model, and that require significant backtracking steps to resolve, the algorithm tends to select more lengthy token sequences that have not yet reached the bar and position of the constraint. This leads to perfectly consistent sequences that are, however, ``overcrowded'' with chords within bars that come before the constraint. These sequences are preferred since they are accumulating more log probability contributions from more tokens. To mitigate this behavior, a penalty is added for overcrowded bars by subtracting the ratio of total tokens over the count of \texttt{<bar>} tokens for each generated sequence. Therefore, the final score for each sequence $S$ is:
\[
\text{score} = \sum_{t \text{ in }S} \text{log} P(t | \text{context}) - \frac{|S|}{|S == \texttt{<bar>}| + 1} \text{,}
\]
where $P(t | \text{context})$ is the probability that the model assigns to a token $t$ in the sequence given the context (melody and previous harmonic tokens) and $|X|$ indicates the number of elements in the set $X$.

In terms of time complexity, the algorithm acts in two stages. During the first stage, the goal is to find a sequence that satisfies the constraints; all possible paths can be potentially explored through backtracking from sequences that do not satisfy the constraints. In this stage, the worst case is equivalent to the Dijkstra algorithm~\citep{dijkstra1959note}, since all beams but one may produce sequences that fail to reach the desired constraint. During the second stage of the algorithm, the constraints have been satisfied and, therefore, the algorithm acts as a beam-search algorithm. Assuming that the constraint occurs at the $C$-th token of the generated sequence, that the vocabulary size is $|V|$ and that the generated sequence has $N$ tokens in total, the time complexity of the first part of the algorithm is $O(|V|^C)$ and of the second $O((N-C)\cdot b \cdot k$). Regarding the first part of the algorithm, the best case is to reach the constraint while in beam search mode and therefore the best case scenario is to have a complexity of $O(C\cdot b \cdot k$). The expected case is a mixture of both, therefore, larger $b$ and $k$ values will expectedly have a negative impact on the number of steps required by the algorithm, with more chances, however, in reaching a better solution more quickly.

\begin{figure}
    \centering
    \includegraphics[width=0.95\linewidth]{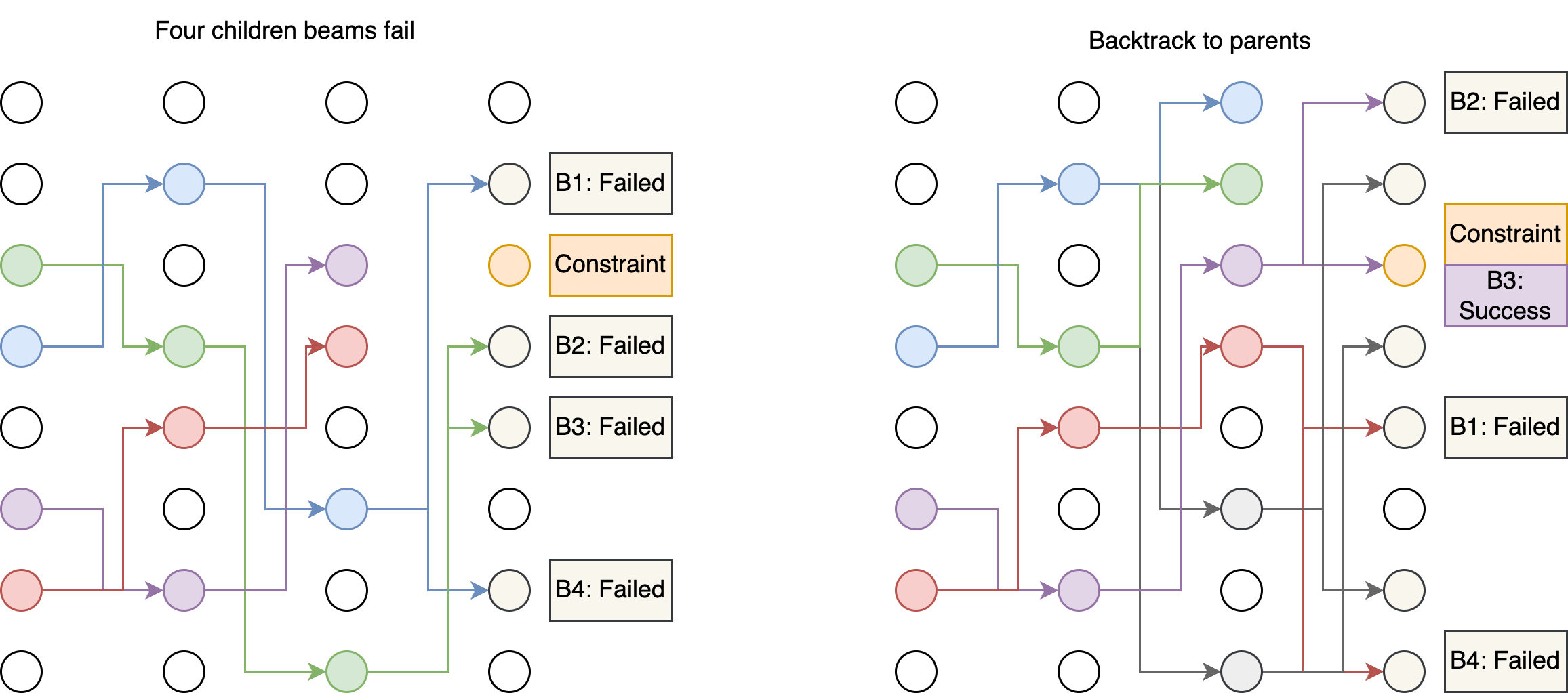}
    \caption{Simplified illustration of two algorithm steps. The left side shows that the children of the blue and green nodes are better rated than the children of the red and purple nodes. However, they fail to reach the constraint and on the next step (right side) the children of the red and purple nodes are tried out, since they are are better scored than the next best children of the red and blue. One of the children of the purple node achieves the constraint and continues from this point on using the typical beam search algorithm.}
    \label{fig:bstar}
\end{figure}

Figure~\ref{fig:bstar} shows a highly simplified image where the green and blue children are better scored than the children of the red and purple and therefore they are examined first (left side). These children fail to satisfy the constraint and the algorithm falls back to the parents. All alternative children beams produced by the green and blue nodes are now of worse quality than the ones produced by the red and purple. One of the children of the purple succeeds in reaching the constraint and therefore, from this point on, this child expands with the typical beam search algorithm. This image is oversimplified mainly because it shows the tokens on a fixed grid, giving the impression that all beams hold the same number of tokens at each step and that they somehow move in parallel. However, this is not the case. Different beams may hold streams of different number of tokens, e.g. having reached different bars with different numbers of chords per bar. Another point of oversimplification is the fact that the constraint is not actually a single token that needs to be achieved, but rather a sequence of tokens (not necessarily consecutive) that indicate the bar, the position and the chord of the constraint.

%% file: 2_results.tex
\section{Results}\label{sec:results}

This section analyzes the extent to which constraints are satisfied, providing also an empirical assessment of the complexity of the algorithm in real-world settings. Additionally, the musical characteristics of the generated pieces are examined using objective metrics that are well-established in the literature. For the remainder of the section the Chord Symbol and Pitch Class tokenization methods will be abbreviated as CS and PC respectively. The employed models include 8 layers with 8 attention heads (in both the encoder and the decoder in BART) with 512 dimension, and achieved a validation set token-level prediction accuracy of around 80\% for the CS models and around 85\% for the PC models during training. Training was performed for 50 epochs and the training loss plateaued at around epoch 40 with no signs of significant overfitting due to high dropout rates (0.3). The application of the B* algorithm involves a beam ($b$) value of 4 and a lookahead ($k$) value of 2. It should be noted here that the algorithm was run with $b$ and $k$ values of 10 \& 5 and 50 \& 25, but the larger the numbers in the setup, the slower the convergence.

\subsection{Constraint satisfaction analysis}\label{subsec:res:constraints}

This section examines to what extent the models are able to satisfy constraints that are selected randomly within pieces of the test set. For each piece in the test set, a random bar is selected and within it a random chord in a random position, which will serve as a constraint in the generative process. Three versions of the GPT2 and BART models are examined:
\begin{description}
    \item[no] constraint tokens were used during training and inference, only the melody and harmony tokens were provided without the addition of the structure and constraints, i.e. no \texttt{'</m>'} token, no subsequent \texttt{'<bar>'}, \texttt{'<fill>'} and no position / chord constraint tokens. Generation is performed ``blindly'' (in terms of constraints) using beam search with 7 beams -- it is ``random'' whether the assumed constraint will be satisfied.
    \item[soft] constraint models, where the structure and constraint tokens are provided during training and inference and the typical beam search algorithm is employed with 7 beams for generation. These versions aim to identify how the incorporation of the structure and constraint tokens will ``persuade'' the model to follow those suggestions during inference.
    \item[B*] models, that are the exact same models trained in the above \textbf{soft} case, but they generate using the proposed B* algorithm. These models will eventually reach the constraints, but a limit is set to 10,000 model calls; if the constraint has not been reached within this limit, we consider that it has failed to reach the constraints.
\end{description}

Counting model calls is a way of assessing the empirical complexity of the algorithm in this specific test set. Model calls counts in the cases of \textbf{no} and \textbf{soft} constraints coincide with the average token length that is generated by each model. These values are given as proxies to provide some estimation about the additional computational (and, subsequently, overall energy) cost of the examined approach in comparison to the typical beam search algorithm. For reference, the average number of tokens in the ground truth dataset using the CS and PC tokenizers are 37.5743 and 67.4349 respectively.

\begin{table}[h]
    \centering
    \begin{tabular}{lcccccccc}
    \toprule
         & & \multicolumn{3}{c}{Constraint success} & & \multicolumn{3}{c}{Avg. model calls} \\
         & & no & soft & B* & & no & soft & B*  \\ \midrule
        BART CS & & 0.2729 & 0.5717 & 0.9480 & & 36.4862 & 37.1434 & 434.8827 \\
        BART PC & & 0.2017 & 0.5257 & 0.9059 & & 62.8243 & 57.1954 & 1925.14016 \\
        GPT2 CS & & 0.2213 & 0.5954 & 0.9618 & & 35.9750 & 35.2086 & 290.4802 \\
        GPT2 PC & & 0.2181 & 0.6289 & 0.9862 & & 62.7355 & 68.7967 & 403.4523 \\
        \bottomrule
    \end{tabular}
    \caption{Constraint satisfaction success when random constraints are assigned in pieces in the test set and average model calls to assess empirically the practical complexity of the algorithm. Model calls in the \textbf{no} and \textbf{soft} models correspond to the number of tokens they generate.}
    \label{tab:success}
\end{table}

Table~\ref{tab:success} shows the success rates and the average number of model calls for all three examined system versions. The success rate of the \textbf{soft} model outperforms the \textbf{no} versions significantly, indicating that simply the inclusion of the structure and constraint tokens during training and during inference, even without the B* algorithm, helps the models identify to some extent the desired constraint. The increase in the percentage is at least twofold and close to threefold, regardless of whether BART or GPT2, or CS or PC tokenizers are employed. The application of the B* for the first 10,000 steps has lead to reaching the constraint in over 90\% of the cases, around 95\% when using the CS tokenizer and over 98\% when using GPT2 with the PC tokenizer.

Regarding average model calls, the column of B* shows the cases that reached a solution that satisfies the constraints and did not reach 10,000 model calls without producing a desired outcome. For each model, the CS tokenizer reaches solutions with fewer steps with the GPT2 being more efficient than BART in all cases. Especially the case of BART with the PC tokenizer appears to be the most inefficient. Regarding the empirical complexity, as computed at the end of Section~\ref{sec:method}, the PC tokenizer has a significantly smaller vocabulary size $|V|$ than the CS tokenizer, since the PC tokenizer can describe all chords with 12 pitch class tokens, while the CS tokenizer employs $12\times 29 = 348$ tokens to describe all 12 root positions of the 29 employed chord qualities. Even though $|V|$ is smaller for PC, the length of the sequences ($N$) that are required to describe the same harmonic content is larger for PC, since it needs to spell out the chords with multiple tokens. This also holds for the constraint position, $C$, which is larger in the sequences tokenized with PC. The fact that the PC variations are empirically slower is theoretically supported by the fact that $C$ (the exponent of the first part of the algorithm) and $N-C$ (the factor that multiplies the constants $b$ and $k$) are expected to have larger values in PC-tokenized sequences.

\subsection{Objective music evaluation}\label{subsec:objective}

Apart from the complexity analysis, this section aims to identify the impact of constraint imposition to the generative process of the involved methods. These experiments incorporate music-relevant evaluation metrics across 1,400 validation pieces. The specific questions of this investigation are:
\begin{enumerate}
\item How effectively do different tokenization methods incorporate the provided chord constraints into generated harmonies?
\item To what extent do user constraints influence musical coherence, harmonic quality, and rhythmic accuracy in generated harmonizations?
\item How do GPT2 and BART differ in their ability to leverage chord constraints to produce musically consistent results?
\end{enumerate}

\subsubsection{Objective music metrics}\label{subsubsec:musical_based_metrics}

We employ three sets of musical-based metrics that evaluate the generated harmonies according to key musical attributes. Specifically, the first two categories for \emph{chord progressions} and \emph{chord-to-melody harmonicity} proposed by~\cite{yeh2021automatic} have been widely used in the literature~\citep{rhyu2022translating, sun2021melody, huang2024emotion} and show how well the chords fit both the harmonic and the melodic context. The final category, \emph{harmonic rhythm}~\citep{wu2024generating}, focuses on the rhythmic placement of chords to ensure that the chord changes align musically with the underlying melody.

\paragraph{Chord Progression Coherence and Diversity} 
\begin{itemize} 
\renewcommand{\labelitemi}{-}
    \item \textbf{Chord Histogram Entropy (CHE)}: CHE measures chord diversity by evaluating the distribution of chord occurrences within a progression. High CHE implies the piece uses a wide variety of distinct chords. Low CHE means a limited harmonic vocabulary.
    \item \textbf{Chord Coverage (CC)}: Counts the distinct chords used throughout a piece. A higher CC indicates richer harmonic vocabulary, enhancing musical interest. 
    \item \textbf{Chord Tonal Distance (CTD)}: Assesses harmonic smoothness by calculating the average tonal distance between consecutive chords. Low CTD means each chord change is tonally close (e.g., moving from C major to G major – a perfect fifth apart), whereas high CTD indicates more distant chord progressions (e.g., C major to F\# major, which are far on the circle of fifths). Thus, a sequence of smooth functional progressions yields a lower average CTD than one with abrupt, distant key changes.
\end{itemize}

\paragraph{Chord/Melody Harmonicity}
\begin{itemize}
\renewcommand{\labelitemi}{-}
    \item \textbf{Chord Tone to non-Chord Tone Ratio (CTnCTR)}: Evaluates harmonic alignment by comparing chord tones versus non-chord melody notes. CTnCTR close to 1.0 means nearly every melody note is part of the accompanying chord (all chord tones), whereas a lower ratio indicates many non-chord tones in the melody.
    \item \textbf{Pitch Consonance Score (PCS)}: PCS evaluates the interval consonance between melody notes and the chord. High PCS means melody–chord intervals are mostly consonant (unison, third, fifth, sixth, etc.), yielding a pleasant blend. Low PCS implies frequent dissonances (seconds, tritones, sevenths).
    \item \textbf{Melody-Chord Tonal Distance (MCTD)}: MCTD quantifies tonal closeness of melody notes to the chord (e.g., how well the melody tone fits in the chord’s key center). Lower values suggest that the chords harmonically align closely with the melody, enhancing overall harmonic quality (the melody note is in the chord or its diatonic scale), whereas high MCTD indicates the melody note is outside the chord’s harmonic context.
\end{itemize}

\paragraph{Harmonic Rhythm Coherence and Diversity}
\begin{itemize}
\renewcommand{\labelitemi}{-}
    \item \textbf{Harmonic Rhythm Histogram Entropy (HRHE)}: Similarly to CHE, it measures the diversity in rhythmic placements of chords in terms of onsets. Higher HRHE means chords occur at many different beat positions with roughly equal frequency – indicating unpredictable or varied harmonic rhythm. Lower HRHE means chords mostly occur at the same beat position each time (e.g., always on the downbeat of each bar), indicating a very regular harmonic rhythm.
    \item \textbf{Harmonic Rhythm Coverage (HRC)}: Number of distinct chord-rhythm patterns. Greater coverage suggests more diverse rhythmic usage throughout the harmonization. If a piece always follows the same pattern (e.g., one chord per bar, every bar), HRC is minimal.
    \item \textbf{Chord Beat Strength (CBS)}: Evaluates the rhythmic placement of chords in relation to strong beats (e.g., downbeats). Lower CBS values indicate consistent placement on strong or metrically important beats, while higher values suggest irregular or syncopated rhythms.
\end{itemize}

In the unconstrained scenario (see Table~\ref{tab:non-constraints}), the PC Tokenizer combined with BART generally produces outputs closest to the ground truth, particularly regarding harmonic and rhythmic metrics like chord diversity and richness along with rhythmic coherence. From a musical perspective, representing chords through pitch classes offers greater flexibility to the model, enabling it to capture subtle harmonic variations and rhythmic nuances that characterize the original musical pieces more closely. However, it is important to note that in terms of chord-to-melody harmonicity metrics (CTnCTR, PCS, MCTD), the GPT2 model paired with the PC Tokenizer achieves the best alignment with the melody. This suggests that the GPT2 model's prompt continuation strategy particularly excels at capturing intricate melodic-chord interactions, producing harmonies that align effectively and consonantly with the given melody.

\begin{table}[ht]
  \centering
  \resizebox{\textwidth}{!}{%
    \begin{tabular}{llccccccccc}
      \toprule
      Model & Tokenizer & CHE & CC & CTD & CTnCTR & PCS & MCTD & HRHE & HRC & CBS \\
      \midrule
       & GT & 1.360 & 4.663 & 0.897 & 0.837 & 0.480 & 1.345 & 0.685 & 2.788 & 0.441 \\
      \midrule
      BART & CS & 0.949 & 2.934 & \textbf{0.892} & 0.802 & 0.443 & 1.397 & 0.403 & 1.890 & 0.276 \\
      BART & PC  & \textbf{0.973} & \textbf{3.021} & 0.924 & 0.788 & 0.424 & 1.415 & \textbf{0.438} & \textbf{1.935} & \textbf{0.317} \\
      GPT2 & CS & 0.887 & 2.741 & 0.911 & 0.759 & 0.384 & 1.449 & 0.420 & 1.838 & 0.286 \\
      GPT2 & PC  & 0.812 & 2.621 & 0.828 & \textbf{0.815} & \textbf{0.454} & \textbf{1.385} & 0.407 & 1.864 & 0.284 \\
      \bottomrule
    \end{tabular}%
  }
  \caption{Objective evaluation (mean values) of tokenizers and models with \textbf{no} chord constraints, using BART and GPT2. The closest values to the Ground Truth (GT) are the better.}
  \label{tab:non-constraints}
\end{table}

However, when \textbf{soft} chord constraints are applied (see Table~\ref{tab:with-constraints}), the CS tokenizer demonstrates notable superiority, especially with the BART model. Explicitly representing chords as holistic musical symbols (root and chord type in one token) allows the model to better respect constraints, thereby producing harmonizations that are more coherent and structurally aligned with musical expectations. Musically, this method likely enables the model to better integrate predefined chords into the generated sequences, maintaining a consistent harmonic language even when forced to include potentially challenging, out-of-context chords. In contrast, the pitch-class-based approach struggles to seamlessly incorporate such constraints, leading to harmonizations that deviate further from ground-truth values in metrics like chord progression smoothness and rhythmic coherence.

\begin{table}[ht]
  \centering
  \resizebox{\textwidth}{!}{%
    \begin{tabular}{llccccccccc}
      \toprule
      Model & Tokenizer & CHE & CC & CTD & CTnCTR & PCS & MCTD & HRHE & HRC & CBS \\
      \midrule
       & GT & 1.359 & 4.663 & 0.897 & 0.837 & 0.480 & 1.345 & 0.685 & 2.789 & 0.441 \\
      \midrule
      BART & CS & \textbf{0.888} & \textbf{2.765} & 0.839 & \textbf{0.805} & \textbf{0.432} & \textbf{1.399} & \textbf{0.432} & \textbf{1.912} & \textbf{0.309} \\
      BART & PC  & 0.772 & 2.507 & 0.825 & 0.788 & 0.403 & 1.427 & 0.279 & 1.560 & 0.184 \\
      GPT2 & CS & 0.839 & 2.650 & \textbf{0.881} & 0.766 & 0.376 & 1.444 & 0.420 & 1.810 & 0.288 \\
      GPT2 & PC  & 0.760 & 2.482 & 0.780 & 0.786 & 0.405 & 1.422 & 0.400 & 1.881 & 0.276 \\
      \bottomrule
    \end{tabular}%
  }
  \caption{Objective evaluation (mean values) of tokenizers and models using the \textbf{soft} constraints approach for different sets of metrics using BART and GPT2. The closest values to the Ground Truth (GT) are the better.}
  \label{tab:with-constraints}
\end{table}

Table~\ref{tab:a_star} shows that the generation with B* further exacerbates the mismatch between generated and ground truth pieces, since all the models are forced to resort to sub-optimal solutions that, however, satisfy the constraints. The superiority of the CS tokenization is retained here as well.

\begin{table}[ht]
  \centering
  \resizebox{\textwidth}{!}{%
    \begin{tabular}{llccccccccc}
      \toprule
      Model & Tokenizer & CHE & CC & CTD & CTnCTR & PCS & MCTD & HRHE & HRC & CBS \\
      \midrule
       & GT & 1.359 & 4.663 & 0.897 & 0.837 & 0.480 & 1.345 & 0.685 & 2.789 & 0.441 \\
      \midrule
      BART & CS & \textbf{0.750} & \textbf{2.489} & \textbf{0.724} & \textbf{0.796} & \textbf{0.421} & 1.403 & 0.131 & 1.304 & 0.108 \\
      BART & PC  & 0.649 & 2.232 & 0.709 & 0.790 & 0.402 & 1.415 & 0.107 & 1.250 & 0.061 \\
      GPT2 & CS & 0.676 & 2.327 & 0.636 & 0.754 & 0.348 & 1.467 & \textbf{0.215} & \textbf{1.553} & \textbf{0.150} \\
      GPT2 & PC  & 0.747 & 2.482 & 0.658 & 0.749 & 0.342 & \textbf{1.471} & 0.172 & 1.381 & 0.114 \\
      \bottomrule
    \end{tabular}%
  }
  \caption{Objective evaluation (mean values) of the \textbf{B*} approach for different sets of metrics using BART and GPT2. The closest values to the Ground Truth (GT) are the better.}
  \label{tab:a_star}
\end{table}

Notably, the models were well-trained (high token-level accuracy), yet they often failed to satisfy the chord constraints when simply prompted (soft models). This indicates the issue is not just undertraining or insufficient capacity, but a more fundamental limitation: standard Transformers, even when trained on data with constraint tokens, may default to musically ‘likely’ continuations rather than adhere to an out-of-context forced chord. In other words, the transformer’s predictive nature can override constraints that contradict with its learned distribution.

%% file: 3_conclusions.tex
\section{Conclusions}\label{sec:conclusions}

Incorporating chord constraints in melodic harmonization with autoregressive generation models has certain peculiarities that necessitate the development of methods with different specifications in comparison to text constraints in text generation methods. The algorithm proposed in this paper, B*, is a brute-force attempt to tackle the specifics of this problem. Even though this algorithm encompasses accelerating characteristics inspired from beam search and A*, its worst case exponential complexity is ultimately prohibitive for real-world applications. The empirical assessment of the complexity in a real dataset showed that in most cases (over 90\% and in some setups over 98\%) the algorithm can generate a harmonic sequence that satisfies the constraints with fewer than 10,000 model calls. Although this limit on the number of model calls is not excessively high for the employed models
(around 50 million parameters each), applying this algorithm to larger models would potentially be useful, albeit prohibitively costly.

The brute-force nature of the algorithm leaves room for significant improvements. One such improvement would be to expand on the benefits of A* by employing more sophisticated heuristics for further accelerating the search. One such heuristic would be to involve a simple trained model, e.g., an LSTM, that helps to evaluate paths as they are built, before they even reach the control of whether they satisfy the constraint. This model could be trained on the inputs and outputs of the transformer model and predict to what extent an input sequence is possible to reach any token in the vocabulary. Input sequences that hold little such possibility for any of the tokens in the constraints can be penalized and move further down the priority list ($Q$).
